\documentclass[10pt]{article}
\usepackage{epsfig}
\usepackage{lineno}
\usepackage{cite}

\newcommand{\be}{\begin{equation}}
\newcommand{\ee}{\end{equation}}
\newcommand{\bea}{\begin{eqnarray}}
\newcommand{\eea}{\end{eqnarray}}

\begin{document}

\title{ \vspace{1cm} Spectroscopic factor of the 1$^{+}$, $^{25}$Al(p,$\gamma$)$^{26}$Si resonance \\ at E$_{x}$= 5.68 MeV}
\author{A. Parikh$^{1,2,} \footnote{anuj.r.parikh@upc.edu}$ , J. Jos\'e$^{1,2}$\\
\\
\small $^1$Departament de F\'isica i Enginyeria Nuclear, \\ \small Universitat
Polit\`ecnica de Catalunya (EUETIB), E-08036 Barcelona, Spain\\
\small $^2$Institut d'Estudis Espacials de Catalunya (IEEC), E-08034 Barcelona,
Spain\\
}
\maketitle

\begin{abstract} Nuclear shell model predictions for the proton spectroscopic factor of the $1^{+}$, E$_{x}$ = 5.68 MeV level in $^{26}$Si are about fifty times smaller than the value suggested by the measured ($\alpha$,$^{3}$He) cross section for the E$_{x}$= 5.69 MeV mirror level in $^{26}$Mg, assuming purely single-particle transfer.  Given that the 5.69 MeV level has been very weakly, if it all, populated in previous studies of the simpler $^{25}$Mg(d,p) reaction, it is unclear if the ($\alpha$,$^{3}$He) result is a true single-particle spectroscopic factor.  If we assume the ($\alpha$,$^{3}$He) result, the thermonuclear rate of the $^{25}$Al(p,$\gamma$)$^{26}$Si reaction would increase by factors of $\approx 6 - 50$ over stellar temperatures of $T \approx 0.05 - 0.2$ GK.  We examine the implications of this enhanced rate for model predictions of nucleosynthesis in classical nova explosions.
\end{abstract}

{\bf PACS number(s):} 21.10.Jx, 26.30.-k, 27.30.+t

\twocolumn

\begin{figure}[t]
\begin{center}
\includegraphics[scale=0.6]{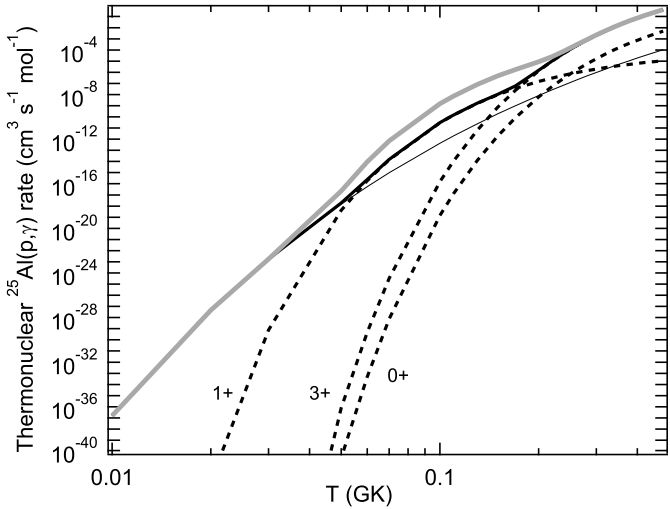}
\caption{The thermonuclear rate of the $^{25}$Al(p,$\gamma$)$^{26}$Si reaction over typical nova temperatures.  Contributions to the total rate of Wrede (2009) (thick black line) by direct capture (thin black line) and the $1^{+}$, $3^{+}$ and $0^{+}$ resonances (dashed lines) are indicated \cite{Wre09}. Also shown is an enhanced rate (grey line) determined using a spectroscopic factor for the $1^{+}$ resonance that is fifty times larger \cite{Yas90} than that used in Ref. \cite{Wre09} (see text).}
\label{fig1}
\end{center}
\end{figure}

Significant progress has been made in recent years to better determine charged-particle thermonuclear reaction rates involved in classical nova explosions.  As a result, most of these rates can be specified with experimentally based uncertainties over the relevant stellar temperatures \cite{Ili10}.   The thermonuclear rate of the $^{25}$Al($p,\gamma$)$^{26}$Si reaction ($Q = 5513.78(48)$ keV \cite{AME12}) is considered to be one of the few remaining rates with an uncertainty large enough to significantly affect model predictions of nucleosynthesis in classical nova explosions (see, e.g., Ref.\cite{Jos11}). Over temperatures involved in novae ($\approx 0.01 - 0.4$ GK) this rate is dominated by contributions from direct capture, a $1^{+}$ resonance at E$_{x}$($^{26}$Si) = 5.68 MeV, and a $3^{+}$ resonance at 5.92 MeV.  As well, a $0^{+}$ resonance at 5.95 MeV is a minor contributor for $T > 0.2$ GK.  These contributions to the total rate, as tabulated in Ref. \cite{Wre09}, are shown in Fig. \ref{fig1}.  We note that recent recommended rates \cite{Wre09, Ili10, Mat10, Ric11} all agree to a factor of $\approx3$ over the relevant temperatures.   

The uncertainty in this rate is difficult to quantify since direct measurements of the strengths of the $1^{+}$ and $3^{+}$ resonances are not yet possible.  Instead, shell model calculations, information from the $\beta$-decay of $^{26}$P, and information from mirror states in $^{26}$Mg have been used to estimate and constrain the proton and $\gamma$-ray partial widths of these two states.  These quantities can then be used to determine resonance strengths and the corresponding resonant contributions of these states to the total thermonuclear rate, as in Fig. \ref{fig1}.  The $1^{+}$, 5.69 MeV and $3^{+}$, 6.13 MeV levels in $^{26}$Mg are the most likely mirror states of the two important resonances in $^{26}$Si due to the lack of other candidates in this energy region of $^{26}$Mg \cite{End98}.  

Recent determinations of the $^{25}$Al($p,\gamma$) rate have used proton spectroscopic factors $S_{l}$ to estimate the required proton partial widths \cite{Wre09, Ili10, Mat10, Ric11}.  These quantities have been calculated using the shell model \cite{Ili96, Mat10, Ric11}, and, for the $3^{+}$ resonance, adopted from the mirror state \cite{End90, Ili96}.  Uncertainties of a factor of three have been assumed for proton partial widths estimated using quantities calculated though the shell model \cite{Bar06,Wre09}.  More recently, Ref. \cite{Ric11} found that a $\approx 20\%$ uncertainty in their theoretical spectroscopic factors accounts for the spread in values determined using different $sd$-shell Hamiltonians.   

Experimental neutron spectroscopic factors determined in the $^{25}$Mg($\alpha$,$^{3}$He)$^{26}$Mg measurement of Yasue et al. (1990) \cite{Yas90} have not been considered, to our knowledge, in any determination of the $^{25}$Al(p,$\gamma$)$^{26}$Si rate.  These measured values are of interest as they are in good agreement with shell model predictions for most of the states populated, including the $3^{+}$ and $0^{+}$ states of concern here.  For example, for the $3^{+}$, $^{26}$Si state at E$_{x}$ = 5.92 MeV, shell model calculations give $S_{l=0} = 0.14$ and $S_{l=2} = 0.33$ \cite{Ili96, Ric11}; Yasue et al. measured $S_{l=0} = 0.14$ and $S_{l=2} = 0.30$ for the mirror state at E$_{x}$($^{26}$Mg) = 6.13 MeV \cite{Yas90}.  For the $0^{+}$, $^{26}$Si state at E$_{x}$ = 5.95 MeV, shell model calculations give $S_{l=2} = 0.047$\cite{Ili96} and $S_{l=2} = 0.039$\cite{Ric11}; Yasue et al. measured $S_{l=2} = 0.054$ for the likely mirror state at E$_{x}$($^{26}$Mg) = 6.26 MeV\cite{Yas90}.  Curiously, for the $1^{+}$, $^{26}$Si state at E$_{x}$ = 5.68 MeV, shell model calculations give $S_{l=2} = 0.0040$\cite{Ili96}, $S_{l=2} = 0.0048$\cite{Mat10}, and $S_{l=2} = 0.0035$\cite{Ric11}; while Yasue et al. measured $S_{l=2} = 0.20$ for the mirror state at E$_{x}$($^{26}$Mg) = 5.69 MeV\cite{Yas90}.  

This large disagreement for the $1^{+}$ resonance may be attributed, in part, to the difficulty in separating the doublet at E$_{x}$($^{26}$Mg) = 5.69 and 5.72 MeV when obtaining angular distributions from the $^{25}$Mg($\alpha$,$^{3}$He) measurement.  This is supported by how differential cross sections for the combined doublet are given at twelve angles between $\theta_{CM} \approx 5 - 60 ^{\circ} $, but cross sections for each member of the doublet are given at only seven angles between $\theta_{CM} \approx 5 - 25 ^{\circ}$ \cite{Yas90}.  A more pressing issue may lie in the suitability of using the ($\alpha$,$^{3}$He) reaction to study the relevant single-particle properties: Yasue et al. themselves suggest that multi-step processes may be responsible for larger ($\alpha$,$^{3}$He) yields for $1^{+}$ states \cite{Yas90}.  Furthermore, the 5.69 MeV state of $^{26}$Mg has been very weakly, if it all, populated in previous studies with a simpler neutron-transfer reaction ($^{25}$Mg(d,p)) and sufficient energy resolution to separate the doublet (e.g., Refs.\cite{Arc84,Bur84}).  As such, it is not clear whether the Yasue et al. result for the $1^{+}$ resonance is a true single-particle spectroscopic factor.  With regard to the thermonuclear rate of the $^{25}$Al(p,$\gamma$) reaction, a spectroscopic factor of $S_{l=2} = 0.20$ for the important $1^{+}$ resonance would lead to a total rate that is $\approx 6 - 50$ times larger than recently recommended rates at T $\approx 0.05 - 0.2$ GK (see Fig. \ref{fig1}). 

New high resolution measurements of the $^{25}$Mg(d,p) and $^{25}$Mg($\alpha$,$^{3}$He) reactions, used in conjunction with improved theoretical treatments of the reaction mechanisms, could help to clarify this possible discrepancy.  To assess the astrophysical motivation for such measurements, we first use some simple calculations to gain some insight into the possible impact in classical nova explosions of the enhanced $^{25}$Al(p,$\gamma$) rate in Fig. \ref{fig1}.  At T = 0.1 GK, the tabulated total rate of Ref. \cite{Wre09} is $3.0 \times 10^{-11}$ cm$^{3}$ s$^{-1}$ mol$^{-1}$, with the resonant contribution of the $1^{+}$ state accounting for $\approx$99\% of this value.  Use of the Yasue et al.\cite{Yas90} spectroscopic factor for the $1^{+}$ resonance would lead to a total rate (at 0.1 GK) of $1.5 \times 10^{-9}$ cm$^{3}$ s$^{-1}$ mol$^{-1}$.  If we then assume a density of $\approx 10^{3}$ g cm$^{-3}$ and a hydrogen mass fraction of $\approx 0.4$ (i.e., after mixing), we would obtain a $^{25}$Al half life for proton capture of $\approx 10^{6}$ s.  Similarly, at T = 0.2 GK, use of the Yasue et al. spectroscopic factor for the $1^{+}$ resonance would lead to a $^{25}$Al half life for proton capture of $\approx 200$ s.  These values may be compared to the half life for the $\beta$-decay of $^{25}$Al: $t_{1/2}$ = 7.2 s.  These rough estimates imply that even the enhanced rate of Fig. \ref{fig1} would not be expected to compete effectively with the $\beta$-decay of $^{25}$Al at the relevant temperatures. 

To investigate this issue in more detail, we performed two hydrodynamic simulations to test the impact of an increased $^{25}$Al(p,$\gamma$) rate on predictions of nucleosynthesis in classical nova explosions.  The two simulations were identical except for the rate of the $^{25}$Al(p,$\gamma$) reaction employed: one used the tabulated reaction rate of Ref. \cite{Wre09}, while the other used an enhanced rate determined using $S_{l=2}=0.20$ for the $1^{+}$ resonance (instead of $S_{l=2} = 0.0040$, as used in Ref. \cite{Wre09}).  These two rates are shown in Fig. \ref{fig1}.  We have adopted a typical nova scenario consisting of a 1.25 M$_{\odot}$ oxygen-neon white dwarf accreting solar material at a rate of $2 \times 10^{-10}$ M$_{\odot}$ yr$^{-1}$ \cite{Jos99}.  Mixing at a level of 50\% between accreted material and the outermost layer of the white dwarf was assumed.    We found no appreciable differences in the nucleosynthesis of the two models; predicted yields were in agreement to better than $\approx5 \%$.  As may be expected from the simple calculations mentioned earlier, even the enhanced $^{25}$Al(p,$\gamma$) rate is not large enough to compete effectively with the $\beta$-decay of any $^{25}$Al produced in nova explosions for T $<0.2$ GK.  For completeness, we performed a third simulation using a $^{25}$Al(p,$\gamma$) rate determined from a spectroscopic factor for the $1^{+}$ resonance that is 50 times lower than the shell model value of $S_{l=2} \approx 0.004$.  Again, as expected, the calculated yields did not differ significantly from those determined using the tabulated rate of Ref. \cite{Wre09}.
 
Nucleosynthesis in the adopted nova model is not very sensitive to variations by a factor of $\approx 50$ of the theoretical proton spectroscopic factor of the $1^{+}$ resonance of the $^{25}$Al(p,$\gamma$)$^{26}$Si reaction.  As such, experimental efforts to improve this rate for models of classical nova explosions should probably focus on better contraining the resonant contribution of the key $3^{+}$ resonance.  

\section*{Acknowledgments}
We thank the anonymous referee for detailed comments that have helped to improve this manuscript.  This work was supported by the Spanish MICINN under Grants No. AYA2010-15685 and No. EUI2009-04167, by the E. U. FEDER funds, and by the ESF EUROCORES Program EuroGENESIS.

%%%%%%%%%         FIG 1       %%%%%%%%%%%%%%%%
%%%%%%%%%%%%  end fig 1   %%%%%%%%%%%%%%%


\begin{thebibliography}{99}
\itemsep -2pt 

\bibitem{Ili10}  C. Iliadis, R. Longland, A.E. Champagne, A. Coc, and R. Fitzgerald, Nucl. Phys. A841, 21 (2010).
\bibitem{AME12} M. Wang, G. Audi, A.H. Wapstra, F.G. Kondev, M. MacCormick, X. Xu, and B. Pfeiffer, Chin. Phys. C36, 1603 (2012).
\bibitem{Jos11} J. Jos\'e and C. Iliadis, Rep. Prog. Phys. 74, 096901 (2011).
\bibitem{Wre09}  C. Wrede, Phys. Rev. C 79, 035803 (2009).
\bibitem{Mat10} A. Matic et al., Phys. Rev. C 82, 025807 (2010).
\bibitem{Ric11} W. A. Richter, B. A. Brown, A. Signoracci, and M. Wiescher, Phys. Rev. C 83, 065803 (2011).
\bibitem{End98} P.M. Endt, Nucl. Phys. A633, 1 (1998). 
\bibitem{End90}  P.M. Endt, Nucl. Phys. A521, 1 (1990).
\bibitem{Ili96}  C. Iliadis, L. Buchmann, P. M. Endt, H. Herndl, and M. Wiescher, Phys. Rev. C 53, 475 (1996).	
\bibitem{Bar06} D.W. Bardayan et al., Phys. Rev. C 74, 045804 (2006).
\bibitem{Yas90}  M. Yasue et al., Phys. Rev. C 42, 1279 (1990).
\bibitem{Bur84} M. Burlein, K.S. Dhuga, and H.T. Fortune, Phys. Rev. C 29, 2013 (1984).
\bibitem{Arc84} H.F.R. Arciszewski, E.A. Bakkum, C.P.M. Van Engelen, P.M.Endt, and R.Kamermans,  Nucl. Phys. A430, 234 (1984).
\bibitem{Jos99} J. Jos\'e, A. Coc,  and M. Hernanz, Astrophys. J. 520, 347 (1999).


\end{thebibliography}
\end{document}